%% file: paretoLaws.tex
\documentclass[12pt]{article}
\usepackage{setspace, natbib, graphicx}
\usepackage[multiple]{footmisc}
\input{mathdef}

\long\def\symbolfootnote[#1]#2{\begingroup
\def\thefootnote{\fnsymbol{footnote}}\footnote[#1]{#2}\endgroup}
\numberwithin{equation}{section}
\newenvironment{proofLemma1}[1][Proof of Lemma \ref{totalWealthLemma}.]{\begin{trivlist}
\item[\hskip \labelsep {\bfseries #1}]}{\hfill\qed\end{trivlist}}
\newenvironment{proofLemma2}[1][Proof of Lemma \ref{localTimeLemma}.]{\begin{trivlist}
\item[\hskip \labelsep {\bfseries #1}]}{\hfill\qed\end{trivlist}}
\newenvironment{proofLemma3}[1][Proof of Lemma \ref{localTimeAKLemma}.]{\begin{trivlist}
\item[\hskip \labelsep {\bfseries #1}]}{\hfill\qed\end{trivlist}}
\newenvironment{proofLemma4}[1][Proof of Lemma \ref{estLocalTimeLemma}.]{\begin{trivlist}
\item[\hskip \labelsep {\bfseries #1}]}{\hfill\qed\end{trivlist}}
\newenvironment{proofTheorem1}[1][Proof of Theorem \ref{wealthDistThm}.]{\begin{trivlist}
\item[\hskip \labelsep {\bfseries #1}]}{\hfill\qed\end{trivlist}}

\theoremstyle{plain}

\renewcommand{\baselinestretch}{1.3}

\begin{document}

\centerline{\bf \LARGE{Empirical Methods for Dynamic Power Law}}

\vskip 15pt

\centerline{\bf \LARGE{Distributions in the Social Sciences}}


\vskip 5pt


\vskip 25pt

\centerline{\large{Ricardo T. Fernholz\symbolfootnote[1]{Robert Day School of Economics and Finance, Claremont McKenna College, 500 E. Ninth St., Claremont, CA 91711, rfernholz@cmc.edu.}}}

\vskip 3pt

\centerline{Claremont McKenna College}

\vskip 25pt


\vskip 5pt

\centerline{\large{\today}}

\vskip 60pt

\renewcommand{\baselinestretch}{1.1}
\begin{abstract}
This paper introduces nonparametric econometric methods that characterize general power law distributions under basic stability conditions. These methods extend the literature on power laws in the social sciences in several directions. First, we show that any stationary distribution in a random growth setting is shaped entirely by two factors---the idiosyncratic volatilities and reversion rates (a measure of cross-sectional mean reversion) for different ranks in the distribution. This result is valid regardless of how growth rates and volatilities vary across different economic agents, and hence applies to Gibrat's law and its extensions. Second, we present techniques to estimate these two factors using panel data. Third, we show how our results offer a structural explanation for a generalized size effect in which higher-ranked processes grow more slowly than lower-ranked processes on average. Finally, we employ our empirical methods using data on commodity prices and show that our techniques accurately describe the empirical distribution of relative commodity prices. We also show the existence of a generalized ``size'' effect for commodities, as predicted by our econometric theory.
\end{abstract}
\renewcommand{\baselinestretch}{1.3}

\vskip 45pt



\vfill

\pagebreak

\section{Introduction} \label{intro}

Power laws are ubiquitous in economics, finance, and the social sciences more broadly. They are found across many different phenomena, ranging from the distribution of income and wealth \citep{Atkinson/Piketty/Saez:2011,Piketty:2014} to the city size distribution \citep{Gabaix:1999} to the distribution of assets of financial intermediaries \citep{janicki/prescott:2006,Fernholz/Koch:2016}. Although a number of potential mechanisms explaining the appearance of power laws have been proposed \citep{Newman:2006,Gabaix:2009}, one of the most broad and influential involves random growth processes.

A large literature in economics, both theoretical and empirical, models different power laws and Pareto distributions as the result of random growth processes that are stabilized by the presence of some friction \citep{Champernowne:1953,Gabaix:1999,Luttmer:2007,Benhabib/Bisin/Zhu:2011}. In this paper, we present rank-based, nonparametric methods that allow for the characterization of general power law distributions in any continuous random growth setting. These techniques, which are well-established and the subject of active research in statistics and mathematical finance, are general and can be applied to Gibrat's law and many of its extensions in economics and finance.\footnote{There is a growing and extensive literature analyzing these rank-based methods. See, for example, \citet{Banner/Fernholz/Karatzas:2005}, \citet{Pal/Pitman:2008}, \citet{Ichiba/Papathanakos/Banner/Karatzas/Fernholz:2011}, and \citet{Shkolnikov:2011}.} According to our general characterization, any stationary distribution in a random growth setting is shaped entirely by two factors---the idiosyncratic volatilities and reversion rates (a measure of cross-sectional mean reversion) for different ranks in the distribution. An increase in idiosyncratic volatilities increases concentration, while an increase in reversion rates decreases concentration. We also present results that allow for the estimation of these two factors using panel data.


Our characterization of a stationary distribution in a general, nonparametric setting provides a framework in which we can understand the shaping forces for almost all power law distributions that emerge in random growth settings. After all, one implication of our results is that the distributional effect of any economic mechanism can be inferred by determining the effect of that mechanism on the idiosyncratic volatilities and reversion rates for different ranked processes. In addition to our general characterization, this is the first paper in economics to provide empirical methods to measure the econometric factors that shape power law distributions using panel data.

These empirical methods allow us to understand the causes, in an econometric sense, of distributional changes that occur for power laws in economics and finance. This econometric analysis has many potential applications. For example, our methods can establish that increasing U.S. income and wealth inequality \citep{Atkinson/Piketty/Saez:2011,Saez/Zucman:2014} is the result of changes in either reversion rates or the magnitude of idiosyncratic shocks to household income and wealth. These econometric changes could then be linked to the evolution of policy, skill-biased technological change, or other changes in the economic environment. This is the approach of \citet{Fernholz/Koch:2016}, who analyze the increasing concentration of U.S. bank assets using the econometric techniques presented in this paper. A similar analysis of increasing U.S. house price dispersion \citep{Nieuwerburgh/Weill:2010} should also yield new conclusions and useful insight.

In order to demonstrate the validity and accuracy of our empirical methods, we estimate reversion rates and idiosyncratic volatilities using monthly commodity prices data from 1980 - 2015 and compare the predicted distribution of relative commodity prices using our methods to the average distribution of relative commodity prices observed during this period. Although our methods apply most naturally to distributions such as wealth, firm size, and city size, they can also be applied to the distributions of relative asset prices. By testing our methods using normalized commodity prices data, we are able to examine the applicability of these methods to relative asset price distributions.

Commodity prices must be normalized so that they can be compared in an economically meaningful way, but as long as these appropriately-normalized prices satisfy the basic regularity conditions that our econometric theory relies on, then our methods should be applicable to the distribution of relative commodity prices. Furthermore, because the distribution of relative normalized commodity prices appears to be stationary during the 1980 - 2015 period, the rank-based reversion rates and idiosyncratic volatilities that we estimate should provide an accurate description of the observed relative commodity price distribution during this period. We confirm that this is in fact the case. One of the contributions of this paper, then, is to show that our empirical methods can validly be applied not only to standard size distributions but also to relative asset price distributions. This result highlights the potential for future applications of our econometric techniques using other data sets.

In addition to our characterization of a stationary distribution in a general random growth setting, we also show that a mean-reversion condition is necessary for the existence of such a stationary distribution. Specifically, a stationary distribution exists only if the growth rates of higher-ranked processes are on average lower than the growth rates of lower-ranked processes. If we let the processes in our general random growth setting represent the total market capitalizations of different stocks, then this mean-reversion condition implies that bigger stocks must generate lower capital gains than smaller stocks. This is similar to the well-known size effect for stocks---the tendency for U.S. stocks with large total market capitalizations to generate lower average returns than U.S. stocks with small total market capitalizations \citep{Banz:1981}.

In terms of normalized commodity prices, the mean-reversion condition that we describe as necessary for the existence of a stationary distribution provides a testable prediction of the existence of a generalized ``size'' effect for commodities. That is, our results predict that higher-ranked, higher-priced, ``bigger'' commodities should generate lower returns on average than lower-ranked, lower-priced, ``smaller'' commodities. Using the same monthly commodity prices data from 1980 - 2015, we confirm that this is in fact the case. We show that an equal-weighted portfolio that invests only in the most expensive (highest ranked) commodities each month generates a yearly return on average more than 6\% below the return on an equal-weighted portfolio that invests only in the least expensive (lowest ranked) commodities each month, exactly as predicted by our econometric theory. Future research that analyzes the risk and liquidity properties of this generalized ``size'' effect for commodities and attempts to determine if this excess return is consistent with standard equilibrium asset pricing theories \citep{Lucas:1978,Fama/French:1993} could yield interesting results.

The rest of this paper is organized as follows. Section \ref{model} presents our nonparametric framework and derives the main result that characterizes general stationary power law distributions. Section \ref{estimation} presents results that show how to estimate the two shaping factors of a power law distribution using panel data. Section \ref{application} presents estimates of rank-based reversion rates and idiosyncratic volatilities using commodity prices data, and also shows the existence of a large generalized ``size'' effect as predicted by our econometric results. Section \ref{conclusion} concludes. Appendix \ref{assumptions} discusses the regularity assumptions needed for our main results, and Appendix \ref{proofs} contains all proofs.

\vskip 70pt

\section{A Nonparametric Approach to Dynamic Power Law Distributions} \label{model}

For consistency, we shall refer to agents holding units throughout this section. However, it is important to note that in this general setup agents can represent households, firms, cities, countries, and other entities, with the corresponding units representing income, wealth, total employees, population, and other quantities. Furthermore, we can also interpret agents' holdings of units as the prices of different assets, as we shall do for commodity prices in Section \ref{application} below.

Consider a population that consists of $N > 1$ agents. Time is continuous and denoted by $t \in \rplu$, and uncertainty in this population is represented by a filtered probability space $(\O, \F, \F_t, P)$. Let $\mathbf{B}(t) = (B_1(t), \ldots, B_M(t))$, $t \in [0, \infty)$, be an $M$-dimensional Brownian motion defined on the probability space, with $M \geq N$. We assume that all stochastic processes are adapted to $\{\F_t; t \in [0, \infty)\}$, the augmented filtration generated by $\mathbf{B}$.\footnote{In order to simplify the exposition, we shall omit many of the less important regularity conditions and technical details involved with continuous-time stochastic processes.}

\subsection{Dynamics}
The total units held by each agent $i = 1, \ldots, N$ is given by the process $x_i$. Each of these unit processes evolves according to the stochastic differential equation
\begin{equation} \label{wealthDynamics}
 d\log x_i(t) = \m_i(t)\,dt + \sum_{s=1}^M\d_{is}(t)\,dB_s(t),
\end{equation}
where $\m_i$ and $\d_{is}$, $s = 1, \ldots, M$, are measurable and adapted processes. The growth rates and volatilities, $\m_i$ and $\d_{is}$, respectively, are general and practically unrestricted, having only to satisfy a few basic regularity conditions that are discussed in Appendix \ref{assumptions}. These conditions imply that the unit processes for the agents are continuous semimartingales, which represent a broad class of stochastic processes (for a detailed discussion, see \citealp{Karatzas/Shreve:1991}).\footnote{This basic setup shares much in common with the continuous-time finance literature (see, for example, \citealp{Karatzas/Shreve:1998,Duffie:2001}). Continuous semimartingales are more general than It{\^o} processes, which are common in the continuous-time finance literature \citep{Nielsen:1999}.}

Indeed, the martingale representation theorem \citep{Nielsen:1999} implies that any plausible continuous process for agents' unit holdings can be written in the nonparametric form of equation \eqref{wealthDynamics}. Furthermore, this section's results can also apply to processes that are subject to sporadic, discontinuous jumps.\footnote{This is an open area for research, but such extensions are examined by \citet{Shkolnikov:2011} and \citet{Fernholz:2016a}.} As a consequence, all previous analyses based on Gibrat's law or specific extensions to Gibrat's law \citep{Gabaix:1999,Gabaix:2009} are special cases of our general framework in this paper.

It is useful to describe the dynamics of the total units held by all agents, which we denote by $x(t) = x_1(t) + \cdots + x_N(t)$. In order to do so, we first characterize the covariance of unit holdings across different agents over time. For all $i, j = 1, \ldots, N$, let the covariance process $\r_{ij}$ be given by
\begin{equation} \label{rhoIJ}
  \r_{ij}(t) = \sum_{s = 1}^M\d_{is}(t)\d_{js}(t).
\end{equation}
Applying \ito's Lemma to equation \eqref{wealthDynamics}, we are now able to describe the dynamics of the total units process $x$.

\begin{lem} \label{totalWealthLemma}
The dynamics of the process for total units held by all agents $x$ are given by
\begin{equation} \label{totalWealthDynamics}
 d\log x(t) = \m(t)\,dt + \sum_{i=1}^N\sum_{s=1}^M\theta_i(t)\d_{is}(t)\,dB_s(t), \as,
\end{equation}
where
\begin{equation} \label{shares}
 \theta_i(t) = \frac{x_i(t)}{x(t)},
\end{equation}
for $i = 1, \ldots, N$, and
\begin{equation} \label{mu}
 \m(t) = \sum_{i=1}^N\theta_i(t)\m_i(t) + \frac{1}{2}\left(\sum_{i=1}^N\theta_i(t)\r_{ii}(t) - \sum_{i,j=1}^N\theta_i(t)\theta_j(t)\r_{ij}(t)\right).
\end{equation}
\end{lem}

\subsection{Rank-Based Dynamics}
In order to characterize the stationary distribution of units in this setup, it is necessary to consider the dynamics of agents' unit holdings by rank. One of the key insights of our approach and of this paper more generally is that rank-based unit dynamics are the essential determinants of the distribution of units. As we demonstrate below, there is a simple, direct, and robust relationship between rank-based unit growth rates and the distribution of units. This relationship is a purely statistical result and hence can be applied to essentially any economic environment, no matter how complex.

The first step in achieving this characterization is to introduce notation for agent rank based on unit holdings. For $k = 1, \ldots, N$, let
\begin{equation} \label{rankWealth}
x_{(k)}(t) = \max_{1 \leq i_1 < \cdots < i_k \leq N} \min \left(x_{i_1}(t), \ldots, x_{i_k}(t)\right),
\end{equation}
so that $x_{(k)}(t)$ represents the units held by the agent with the $k$-th most units among all the agents in the population at time $t$. For brevity, we shall refer to this agent as the $k$-th largest agent throughout this paper. One consequence of this definition is that
\begin{equation}
 \max (x_1(t), \ldots, x_N(t)) = x_{(1)}(t) \geq x_{(2)}(t) \geq \cdots \geq x_{(N)}(t) = \min (x_1, \ldots, x_N(t)).
\end{equation}
Next, let $\theta_{(k)}(t)$ be the share of total units held by the $k$-th largest agent at time $t$, so that
\begin{equation} \label{rankShares}
\theta_{(k)}(t) = \frac{x_{(k)}(t)}{x(t)},
\end{equation}
for $k = 1, \ldots, N$. 

The next step is to describe the dynamics of the agent rank unit processes $x_{(k)}$ and rank unit share processes $\theta_{(k)}$, $k = 1, \ldots, N$. Unfortunately, this task is complicated by the fact that the max and min functions from equation \eqref{rankWealth} are not differentiable, and hence we cannot simply apply \ito's Lemma in this case. Instead, we introduce the notion of a local time to solve this problem. For any continuous process $z$, the \emph{local time} at $0$ for $z$ is the process $\L_z$ defined by
\begin{equation} \label{localTime}
 \L_z(t) = \frac{1}{2}\left(|z(t)| - |z(0)| - \intt\sgn(z(s))\,dz(s)\right).
\end{equation}
As detailed by \citet{Karatzas/Shreve:1991}, the local time for $z$ measures the amount of time the process $z$ spends near zero.\footnote{For more discussion of local times, and especially their connection to rank processes, see \citet{Fernholz:2002}.} To be able to link agent rank to agent index, let $p_t$ be the random permutation of $\{1, \ldots, N\}$ such that for $1 \leq i, k \leq N$,
\begin{equation} \label{pTK}
 p_t(k) = i \quad \text{if} \quad x_{(k)}(t) = x_i(t).
\end{equation}
This definition implies that $p_t(k) = i$ whenever agent $i$ is the $k$-th largest agent in the population at time $t$, with ties broken in some consistent manner.\footnote{For example, if $x_i(t) = x_j(t)$ and $i > j$, then we can set $p_t(k) = i$ and $p_t(k+1) = j$.}

\begin{lem} \label{localTimeLemma}
For all $k = 1, \ldots, N$, the dynamics of the agent rank unit processes $x_{(k)}$ and rank unit share processes $\theta_{(k)}$ are given by
\begin{equation} \label{rankWealthDynamics}
 d\log x_{(k)}(t) = d\log x_{p_t(k)}(t) + \frac{1}{2}d\L_{\log x_{(k)} - \log x_{(k + 1)}}(t) - \frac{1}{2}d\L_{\log x_{(k - 1)} - \log x_{(k)}}(t),
\end{equation}
a.s, and
\begin{equation} \label{rankWealthShareDynamics1}
  d\log\theta_{(k)}(t) = d\log\theta_{p_t(k)}(t) + \frac{1}{2}d\L_{\log\theta_{(k)} - \log\theta_{(k + 1)}}(t) - \frac{1}{2}d\L_{\log\theta_{(k - 1)} - \log\theta_{(k)}}(t),
\end{equation}
a.s., with the convention that $\L_{\log x_{(0)} - \log x_{(1)}}(t) = \L_{\log x_{(N)} - \log x_{(N+1)}}(t) = 0$.
\end{lem}

According to equation \eqref{rankWealthDynamics} from the lemma, the dynamics of units for the $k$-th largest agent in the population are the same as those for the agent that is the $k$-th largest at time $t$ (agent $i = p_t(k)$), plus two local time processes that capture changes in agent rank (one agent overtakes another in unit holdings) over time.\footnote{For brevity, we write $dz_{p_t(k)}(t)$ to refer to the process $\sum_{i = 1}^N1_{\{i = p_t(k)\}}dz_i(t)$ throughout this paper.} Equation \eqref{rankWealthShareDynamics1} describes the similar dynamics of the rank unit share processes $\theta_{(k)}$.


Using equations \eqref{wealthDynamics} and \eqref{totalWealthDynamics} and the definition of $\theta_i(t)$, we have that for all $i = 1, \ldots, N$,
\begin{align}
 d\log\theta_i(t) & = d\log x_i(t) - d\log x(t) \notag \\
 & = \m_i(t)\,dt + \sum_{s=1}^M\d_{is}(t)\,dB_s(t) - \m(t)\,dt - \sum_{i=1}^N\sum_{s=1}^M\theta_i(t)\d_{is}(t)\,dB_s(t). \label{wealthShareDynamics}
\end{align}
If we apply Lemma \ref{localTimeLemma} to equation \eqref{wealthShareDynamics}, then it follows that
\begin{equation} \label{rankWealthShareDynamics2}
\begin{aligned}
 d\log\theta_{(k)}(t) & = \left(\m_{p_t(k)}(t) - \m(t)\right)\,dt + \sum_{s=1}^M\d_{p_t(k)s}(t)\,dB_s(t) - \sum_{i=1}^N\sum_{s=1}^M\theta_i(t)\d_{is}(t)\,dB_s(t) \\
 & \qquad + \frac{1}{2}d\L_{\log\theta_{(k)} - \log\theta_{(k + 1)}}(t) - \frac{1}{2}d\L_{\log\theta_{(k - 1)} - \log\theta_{(k)}}(t),
\end{aligned}
\end{equation}
a.s, for all $k = 1, \ldots, N$. Equation \eqref{rankWealthShareDynamics2}, in turn, implies that the process $\log\theta_{(k)} - \log\theta_{(k+1)}$ satisfies, a.s., for all $k = 1, \ldots, N - 1$,
\begin{equation} \label{rankWealthShareDynamics3}
\begin{aligned}
d\left(\log\theta_{(k)}(t) - \log\theta_{(k+1)}(t)\right) & =  \left(\m_{p_t(k)}(t) - \m_{p_t(k+1)}(t)\right)\,dt + d\L_{\log\theta_{(k)} - \log\theta_{(k + 1)}}(t) \\
& \qquad - \frac{1}{2}d\L_{\log\theta_{(k-1)} - \log\theta_{(k)}}(t) - \frac{1}{2}d\L_{\log\theta_{(k+1)} - \log\theta_{(k + 2)}}(t) \\
& \qquad + \sum_{s=1}^M\left(\d_{p_t(k)s}(t) - \d_{p_t(k+1)s}(t)\right)\,dB_s(t).
\end{aligned}
\end{equation}
The processes for relative unit holdings of adjacent agents in the distribution of units as given by equation \eqref{rankWealthShareDynamics3} are key to describing the distribution of units in this setup.

\subsection{Stationary Distribution}
The results presented above allow us to analytically characterize the stationary distribution of units in this setup. Let $\a_k$ equal the time-averaged limit of the expected growth rate of units for the $k$-th largest agent relative to the expected growth rate of units for the entire population of agents, so that
\begin{equation} \label{alphaK}
 \a_k = \limT1\intT\left(\m_{p_t(k)}(t) - \m(t)\right)\,dt,
\end{equation}
for $k = 1, \ldots, N$. The relative growth rates $\a_k$ are a rough measure of the rate at which agents' unit holdings revert to the mean. We shall refer to the $-\a_k$ as reversion rates, since lower values of $\a_k$ (and hence higher values of $-\a_k$) imply faster cross-sectional mean reversion.


In a similar manner, we wish to define the time-averaged limit of the volatility of the process $\log\theta_{(k)} - \log\theta_{(k + 1)}$, which measures the relative unit holdings of adjacent agents in the distribution of units. For all $k = 1, \ldots, N - 1$, let $\s_k$ be given by
\begin{equation} \label{sigmaK}
 \s^2_k = \limT1\intT\sum_{s=1}^M\left(\d_{p_t(k)s}(t) - \d_{p_t(k+1)s}(t)\right)^2\,dt.
\end{equation}
The relative growth rates $\a_k$ together with the volatilities $\s_k$ entirely determine the shape of the stationary distribution of units in this population, as we shall demonstrate below.

We shall refer to the volatility parameters $\s_k$, which measure the standard deviations of the processes $\log \theta_{(k)} - \log \theta_{(k + 1)}$, as idiosyncratic volatilities. An idiosyncratic shock to the unit holdings of either the $k$-th or $(k+1)$-th ranked agent alters the value of $\log \theta_{(k)} - \log \theta_{(k + 1)}$ and hence will be measured by $\s_k$. In addition, however, a shock that affects the unit holdings of multiple agents that do not occupy adjacent ranks in the distribution will also alter this value. Indeed, any shock that affects $\log \theta_{(k)}$ and $\log \theta_{(k + 1)}$ differently, must necessarily alter the value of $\log \theta_{(k)} - \log \theta_{(k + 1)}$ and hence will be measured by $\s_k$. In this sense, the volatility parameters $\s_k$ are slightly more general than pure idiosyncratic volatilities that capture only shocks that affect one single agent at a time.

Finally, for all $k = 1, \ldots, N$, let
\begin{equation} \label{kappa}
 \k_k = \limT1\L_{\log\theta_{(k)} - \log\theta_{(k + 1)}}(T).
\end{equation}
Let $\k_0 = 0$, as well. Throughout this paper, we assume that the limits in equations \eqref{alphaK}-\eqref{kappa} do in fact exist. In Appendix \ref{proofs}, we show that the parameters $\a_k$ and $\k_k$ are related by $\a_k - \a_{k+1} = \frac{1}{2}\k_{k-1} - \k_k + \frac{1}{2}\k_{k+1}$, for all $k = 1, \ldots, N-1$.


The \emph{stable version} of the process $\log\theta_{(k)} - \log\theta_{(k + 1)}$ is the process $\log\hat{\theta}_{(k)} - \log\hat{\theta}_{(k + 1)}$ defined by
\begin{equation} \label{stableVersion}
  d\left(\log\hat{\theta}_{(k)}(t) - \log\hat{\theta}_{(k+1)}(t)\right) = -\k_k\,dt + d\L_{\log\hat{\theta}_{(k)} - \log\hat{\theta}_{(k + 1)}}(t) + \s_k\,dB(t),
\end{equation}
for all $k = 1, \ldots, N - 1$.\footnote{For each $k = 1, \ldots, N-1$, equation \eqref{stableVersion} implicitly defines another Brownian motion $B(t)$, $t \in [0, \infty)$. These Brownian motions can covary in any way across different $k$.} The stable version of $\log\theta_{(k)} - \log\theta_{(k+1)}$ replaces all of the processes from the right-hand side of equation \eqref{rankWealthShareDynamics3} with their time-averaged limits, with the exception of the local time process $\L_{\log\theta_{(k)} - \log\theta_{(k + 1)}}$. By considering the stable version of these relative unit holdings processes, we are able to obtain a simple characterization of the distribution of units.


\begin{thm} \label{wealthDistThm}
There is a stationary distribution for the stable version of unit holdings by agents in this population if and only if $\a_1 + \cdots + \a_k < 0$, for $k = 1, \ldots, N - 1$. Furthermore, if there is a stationary distribution of units, then for $k = 1, \ldots, N - 1$, this distribution satisfies
\begin{equation} \label{wealthDistEq}
 E\left[\log\hat{\theta}_{(k)}(t) - \log\hat{\theta}_{(k + 1)}(t)\right] = \frac{\s^2_k}{-4(\a_1 + \cdots + \a_k)}, \as\end{equation}
\end{thm}

Theorem \ref{wealthDistThm} provides an analytic rank-by-rank characterization of the entire distribution of units. This is achieved despite minimal assumptions on the processes that describe the dynamics of agents' unit holdings over time. As long as the relative growth rates, volatilities, and local times that we take limits of in equations \eqref{alphaK}-\eqref{kappa} do not change drastically and frequently over time, then the distribution of the stable versions of $\theta_{(k)}$ from Theorem \ref{wealthDistThm} will accurately reflect the distribution of the true versions of these rank unit share processes.\footnote{\citet{Fernholz:2002} and \citet{Fernholz/Koch:2016} demonstrate the accuracy of Theorem \ref{wealthDistThm} in matching, respectively, the distribution of total market capitalizations of U.S. stocks and the distribution of assets of U.S. financial intermediaries.} For this reason, we shall assume that equation \eqref{wealthDistEq} approximately describes the true versions of $\theta_{(k)}$ throughout much of this paper.



The theorem yields two important insights. First, it shows that an understanding of rank-based unit holdings dynamics is sufficient to describe the entire distribution of units. It is not necessary to directly model and estimate agents' unit holdings dynamics by name, denoted by index $i$, as is common in the literatures on income and wealth inequality \citep{Guvenen:2009,Benhabib/Bisin/Zhu:2011,Altonji/Smith/Vidangos:2013}. Second, the theorem shows that the only two factors that affect the distribution of units are the rank-based reversion rates, $-\a_k$, and the rank-based volatilities, $\s_k$.

The characterization in equation \eqref{wealthDistEq} is flexible enough to replicate any empirical distribution. Indeed, regardless of whether the true distribution of units is Pareto, log-normal, double Pareto log-normal, or something else, Theorem \ref{wealthDistThm} implies that this distribution appears asymptotically for certain values of the reversion rates and volatilities.

According to Theorem \ref{wealthDistThm}, stationarity of the distribution of agents' unit holdings requires that the reversion rates $-\a_k$ must sum to positive quantities, for all $k = 1, \ldots, N - 1$. Stability, then, requires a mean reversion condition in the sense that the growth rate of units for the agents with the most units in the population must be strictly below the growth rate of units for agents with smaller unit holdings. The unstable case in which this mean reversion condition does not hold is examined in detail by \citet{Fernholz/Fernholz:2014} and \citet{Fernholz:2016b}. As we shall demonstrate in Section \ref{application}, this condition has significant implications for the dynamics of different ranked commodity prices.

If we impose more restrictions on the stable versions of the relative unit holdings processes $\log\hat{\theta}_{(k)} - \log\hat{\theta}_{(k + 1)}$, then it is possible to link the reversion rates $-\a_k$ and volatilities $\s_k$ to mobility. In particular, if we assume that agents face only aggregate and idiosyncratic shocks to their unit holdings, then it is possible to show that mobility is increasing in cross-sectional mean reversion $-\a_k$ and decreasing in unit concentration, as measured by the expected value of $\log\hat{\theta}_{(k)} - \log\hat{\theta}_{(k + 1)}$. Mobility in this context is measured as the expected time for $x_{(k+1)}$ to overtake the higher ranked $x_{(k)}$. A proof of this result and some extensions can be found in \citet{Fernholz:2016a}.

\subsection{Gibrat's Law, Zipf's Law, and Pareto Distributions} \label{gibrat}
It is useful to see how our rank-based, nonparametric approach nests many common examples of random growth processes from other literatures as special cases. We shall focus on the influential example of Gibrat's law, and also describe the conditions that are necessary for Gibrat's law to give rise to Zipf's law.

According to \citet{Gabaix:2009}, the strongest form of Gibrat's law for unit holdings imposes growth rates and volatilities that do not vary across the distribution of unit holdings. In terms of the reversion rates $-\a_k$ (which measure relative unit growth rates for different ranked agents) and idiosyncratic volatilities $\s_k$, this requirement is equivalent to there existing some common $\a < 0$ and $\s > 0$ such that
\begin{equation} \label{gibratAlpha}
 \a = \a_1 = \cdots = \a_{N-1},
\end{equation}
and
\begin{equation} \label{gibratSigma}
 \s = \s_1 = \cdots = \s_{N-1}.
\end{equation}
In terms of equation \eqref{wealthDistEq} from Theorem \ref{wealthDistThm}, then, Gibrat's law yields unit shares that satisfy
\begin{equation} \label{gibratSS}
 E\left[\log\hat{\theta}_{(k)}(t) - \log\hat{\theta}_{(k + 1)}(t)\right] = \frac{\s^2_k}{-4(\a_1 + \cdots + \a_k)} = \frac{\s^2}{-4k\a} \as,
\end{equation}
for all $k = 1, \ldots, N - 1$.

The distribution of agents' unit holdings follows a Pareto distribution if a plot of unit shares as a function of rank, using log scales for both axes, appears as a straight line.\footnote{See the discussions in \citet{Newman:2006} and \citet{Gabaix:2009}.} Furthermore, if the slope of such a straight line plot is -1, then agents' unit shares obey Zipf's law \citep{Gabaix:1999}. According to equation \eqref{gibratSS}, for all $k = 1, \ldots, N - 1$, the slope of such a log-log plot in the case of Gibrat's law is given by
\begin{equation} \label{gibratPareto}
 \frac{E\left[\log\hat{\theta}_{(k)}(t) - \log\hat{\theta}_{(k + 1)}(t)\right]}{\log k - \log k+1} \approx -kE\left[\log\hat{\theta}_{(k)}(t) - \log\hat{\theta}_{(k + 1)}(t)\right] = \frac{-k\s^2}{-4k\a}  = \frac{\s^2}{4\a}.
\end{equation}
Equation \eqref{gibratPareto} shows that Gibrat's law yields a Pareto distribution in which the log-log plot of unit shares versus rank has slope $\s^2/4\a < 0$, which is equivalent to the Pareto distribution having parameter $-\s^2/4\a > 0$. Furthermore, we see that agents' unit shares obey Zipf's law only if $\s^2 = -4\a$, in which case the log-log plot has slope -1.

Theorem \ref{wealthDistThm} thus demonstrates that Gibrat's law and Zipf's law are special cases of general power law distributions in which growth rates and volatilities potentially vary across different ranks in the distribution of unit holdings. Indeed, equation \eqref{wealthDistEq} implies that any power law exponent can obtain in any part of the distribution curve. This flexibility is a novel feature of our empirical methodology and is necessary to accurately match many empirical distributions. For example, \citet{Fernholz/Koch:2016} find that asset growth rates and volatilities vary substantially across different size-ranked U.S. financial intermediaries. Similarly, \citet{Fernholz:2002} finds that growth rates and volatilities of total market capitalization vary substantially across different size-ranked U.S. stocks, while \citet{Neumark/Wall/Zhang:2011} find that employment growth rates vary across different size-ranked U.S. firms. In Section \ref{application}, we confirm this general pattern and show that the growth rates of commodity prices also differ across ranks in a statistically significant and economically meaningful way.


\vskip 70pt

\section{Estimation} \label{estimation}

In order to estimate the reversion rates and volatilities from equation \eqref{wealthDistEq} from Theorem \ref{wealthDistThm}, we use discrete-time approximations of the continuous processes that yield the theorem. For the estimation of the volatility parameters $\s^2_k$, we use the discrete-time approximation of equation \eqref{sigmaK} above. In particular, these estimates are given by
\begin{equation} \label{estSigmaK}
 \s^2_k = \frac{1}{T}\sum_{t = 1}^{T}\left[\left(\log\theta_{p_t(k)}(t + 1) - \log\theta_{p_t(k + 1)}(t + 1)\right) - \left(\log\theta_{p_t(k)}(t) - \log\theta_{p_t(k + 1)}(t)\right)\right]^2,
\end{equation}
for all $k = 1, \ldots, N-1$. Note that $T$ is the total number of periods covered in the data.

The estimation of the rank-based relative growth rates $\a_k$ is more difficult. In order to estimate these parameters, we first estimate the local time parameters $\k_k$ and then exploit the relationship that exists between these local times and the rank-based relative growth rates.

\begin{lem} \label{localTimeAKLemma}
The relative growth rate parameters $\a_k$ and the local time parameters $\k_k$ satisfy
\begin{equation} \label{alphaKappa}
 \a_k = \frac{1}{2}\k_{k-1} - \frac{1}{2}\k_k,
\end{equation}
for all $k = 1, \ldots, N - 1$, and $\a_N = -(\a_1 + \cdots + \a_{N - 1})$.
\end{lem}

\begin{lem} \label{estLocalTimeLemma}
The ranked agent unit share processes $\theta_{(k)}$ satisfy the stochastic differential equation
\begin{equation} \label{estLocalTime1}
\begin{aligned}
 d\log\left(\theta_{p_t(1)}(t) + \cdots + \theta_{p_t(k)}(t)\right) & = d\log \left(\theta_{(1)}(t) + \cdots + \theta_{(k)}(t)\right) \\
  & \qquad - \frac{\theta_{(k)}(t)}{2(\theta_{(1)}(t) + \cdots + \theta_{(k)}(t))}d\L_{\log\theta_{(k)} - \log\theta_{(k + 1)}}(t), \as,
\end{aligned}
\end{equation}
for all $k = 1, \ldots, N$.
\end{lem}

These lemmas together allow us to generate estimates of the rank-based relative growth rates $\a_k$. In order to accomplish this, we first estimate the local time processes $\L_{\log\hat{\theta}_{(k)} - \log\hat{\theta}_{(k + 1)}}$ using the discrete-time approximation of equation \eqref{estLocalTime1}. This discrete-time approximation implies that for all $k = 1, \ldots, N$,
\begin{equation} \label{estLocalTime2}
\begin{aligned}
 \log & \left(\theta_{p_t(1)}(t + 1) + \cdots + \theta_{p_t(k)}(t + 1)\right) - \log\left(\theta_{p_t(1)}(t) + \cdots + \theta_{p_t(k)}(t)\right) = \\
 & \qquad \log \left(\theta_{p_{t+1}(1)}(t + 1) + \cdots + \theta_{p_{t + 1}(k)}(t + 1)\right) - \log \left(\theta_{p_t(1)}(t) + \cdots + \theta_{p_t(k)}(t)\right) \\
 & \qquad \qquad - \frac{\theta_{p_t(k)}(t)}{2\left(\theta_{p_t(1)}(t) + \cdots + \theta_{p_t(k)}(t)\right)}\left(\L_{\log\theta_{(k)} - \log\theta_{(k + 1)}}(t + 1) - \L_{\log\theta_{(k)} - \log\theta_{(k + 1)}}(t)\right),
\end{aligned}
\end{equation}
which, after simplification and rearrangement, yields
\begin{equation} \label{estLocalTime3}
\begin{aligned}
 \L_{\log\theta_{(k)} - \log\theta_{(k + 1)}}(t + 1) & - \L_{\log\theta_{(k)} - \log\theta_{(k + 1)}}(t) = \bigg[\log \left(\theta_{p_{t+1}(1)}(t + 1) + \cdots + \theta_{p_{t + 1}(k)}(t + 1)\right) \\
  & \quad - \log \left(\theta_{p_t(1)}(t + 1) + \cdots + \theta_{p_t(k)}(t + 1)\right)\bigg] \frac{2\left(\theta_{p_t(1)}(t) + \cdots + \theta_{p_t(k)}(t)\right)}{\theta_{p_t(k)}(t)}.
\end{aligned}
\end{equation}
As with our estimates of the volatility parameters $\s^2_k$, we estimate the values of the local times in equation \eqref{estLocalTime3} for $t = 1, \ldots, T$, where $T$ is the total number of periods covered in the data. We also set $\L_{\log\theta_{(k)} - \log\theta_{(k + 1)}}(0) = 0$, for all $k = 1, \ldots, N$.

After estimating the local times in equation \eqref{estLocalTime3}, we then use equation \eqref{kappa} to generate estimates of $\k_k$ according to
\begin{equation} \label{estKappa}
 \k_k = \frac{1}{T}\L_{\log\theta_{(k)} - \log\theta_{(k + 1)}}(T),
\end{equation}
for all $k = 1, \ldots, N$. Finally, we can use the relationship between the parameters $\a_k$ and $\k_k$ established by Lemma \ref{localTimeAKLemma}. This is accomplished via equation \eqref{alphaKappa}, which yields estimates of each $\a_k$ using our estimates of the parameters $\k_k$ from equation \eqref{estKappa}.

While the methods described in this section explain how to generate point estimates of the reversion rates $-\a_k$ and idiosyncratic volatilities $\s_k$, it is important to also understand how much variation there is in these estimates. It is not possible to generate confidence intervals using classical techniques in this setting because the empirical distribution of the parameters $\a_k$ and $\s_k$ is unknown. However, it is possible to use bootstrap resampling to generate confidence intervals for these estimated factors.

Equations \eqref{estSigmaK} and \eqref{estLocalTime3} show that the reversion rates $-\a_k$ and idiosyncratic volatilities $\s_k$ are measured as changes from one period, $t$, to the next, $t+1$. As a consequence, the bootstrap resamples we construct consist of $T-1$ pairs of observations of agents' unit holdings from adjacent time periods (periods $t$ and $t+1$). Such resamples, of course, are equivalent to the full sample which has observations over $T$ periods and hence consists of $T-1$ pairs of observations from adjacent periods. The confidence intervals are then generated by determining the range of values that obtain for the parameters $\a_k$ and $\s_k$ over all of the bootstrap resamples. In Section \ref{application}, we apply our techniques to the distribution of relative commodity prices and generate confidence intervals for our estimates of the reversion rates $-\a_k$ and idiosyncratic volatilities $\s_k$ following this procedure.


\vskip 70pt

\section{Application: The Distribution of Commodity Prices} \label{application}

We wish to confirm the validity and accuracy of the empirical methods we presented in Section \ref{model}. We do this using a publicly available data set on the global monthly spot prices of 22 common commodities for 1980 - 2015 obtained from the Federal Reserve Bank of St. Louis (FRED).\footnote{These commodities are aluminum, bananas, barley, beef, Brent crude oil, cocoa, copper, corn, cotton, iron, lamb, lead, nickel, orange, poultry, rubber, soybeans, sugar, tin, wheat, wool (fine), and zinc.}

In order to accomplish this, we shall use the results and procedure described in Section \ref{estimation} to estimate rank-based reversion rates $-\a_k$ and idiosyncratic volatilities $\s_k$ for the distribution of relative commodity prices over our sample period 1980 - 2015. In this section, then, we shall interpret agents' holdings of units $x_i(t)$ from equation \eqref{wealthDynamics} as the prices of different commodities.\footnote{As long as commodity prices satisfy the basic regularity conditions of Appendix \ref{assumptions} and the distribution of relative commodity prices is stationary, then the econometric results of Sections \ref{model} and \ref{estimation} can be applied.} Because commodities are sold in different units and hence their prices cannot be compared in an economically meaningful way, it is important to normalize these prices by equalizing them in the initial period.

\begin{figure}[ht]
\begin{center}
\vspace{-30pt}
\hspace{-15pt}\scalebox{1.66}{ {\includegraphics[width=0.5\textwidth]{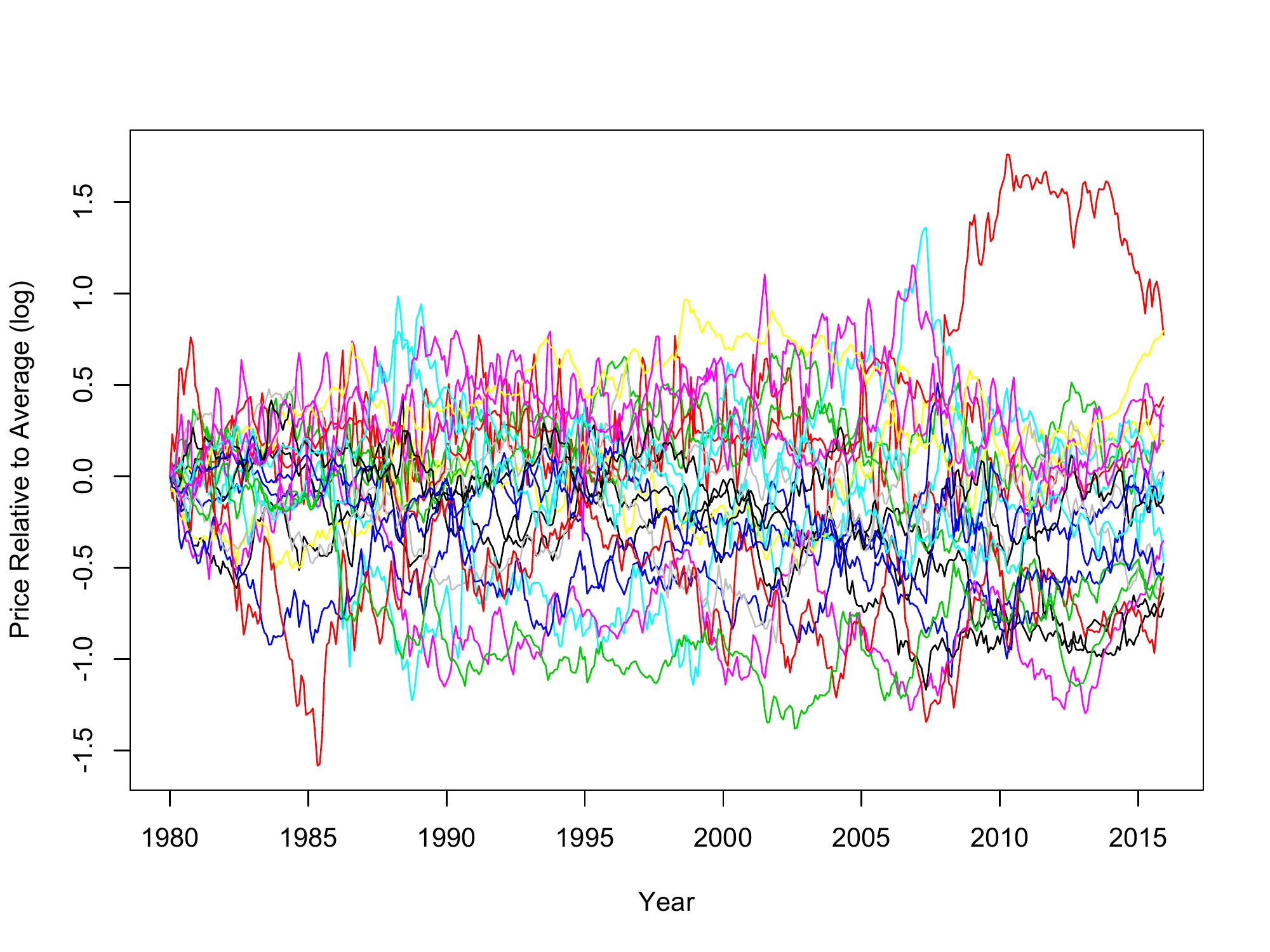}}}
\end{center}
\vspace{-24pt} \caption{Log prices of commodities relative to the average price of all commodities, 1980 - 2015.}
\label{pricesFig}
\end{figure}

The results of Sections \ref{model} and \ref{estimation} apply to the distribution of the parameters $\theta_{(k)}$, $k = 1, \ldots, N$, which in those sections represented the shares of total units held by different ranked agents. If we interpret the $x_i$ as commodity prices, then the parameters $\theta_{(k)}$ represent  commodity ``price shares,'' a quantity that is well defined but difficult to interpret economically. It is easy to show, however, that the distribution of these commodity ``price shares'' $\theta_{(k)}$ is the same as the distribution of commodity prices relative to the average of all commodity prices. This latter quantity has a clear economic interpretation. In this section, we estimate reversion rates and idiosyncratic volatilities that describe the stationary distribution of relative normalized commodity prices according to equation \eqref{wealthDistEq}. As Figure \ref{pricesFig} demonstrates, the distribution of these relative normalized prices appears to be roughly stationary over time. Consistent with this observation, we confirm below that the methods presented in Sections \ref{model} and \ref{estimation} do in fact accurately describe this stationary distribution.

If we let $\bar{x}(t)$ equal the average price of all $N$ commodities at time $t$, then for all $i = 1, \ldots, N$, the relative price of commodity $i$ at time $t$ is defined as
\begin{equation}
 \tilde{x}_i(t) = \frac{x_i(t)}{\bar{x}(t)} = \frac{x_i(t)}{\frac{x_1(t) + \cdots + x_N(t)}{N}} =  \frac{Nx_i(t)}{x_1(t) + \cdots + x_N(t)}.
\end{equation}
The relative price $\tilde{x}_i(t)$ is equal to the price of commodity $i$ at time $t$ relative to the average price of all $N$ commodities at time $t$. If we let $\tilde{x}_{(k)}(t)$ denote the relative price of the $k$-th ranked commodity at time $t$, then  equations \eqref{shares} and \eqref{rankShares} imply that, for all $i, k = 1, \ldots, N$,
\begin{equation} \label{rankedComm}
 \tilde{x}_i(t) = N\theta_i(t) \qquad \text{and} \qquad \tilde{x}_{(k)}(t) = N\theta_{(k)}(t).
\end{equation}
Note that the $k$-th ranked commodity at time $t$ refers to the commodity with the $k$-th highest price at time $t$. It follows from equation \eqref{rankedComm} that for all $k = 1, \ldots, N-1$ and all $t$,
\begin{equation}
 \log\tilde{x}_{(k)}(t) -  \log\tilde{x}_{(k+1)}(t) =  \log\theta_{(k)}(t) -  \log\theta_{(k+1)}(t),
\end{equation}
and hence equation \eqref{wealthDistEq} from Theorem \ref{wealthDistThm} describes both the distribution of commodity ``price shares,'' $\theta_{(k)}$, and relative commodity prices, $\tilde{x}_{(k)}$. In other words, all of our previous results apply to the distribution of relative commodity prices as well.

\subsection{Prediction and Data}
The econometric results of Section \ref{model} suggest that any stationary size distribution can be accurately characterized by the reversion rates $-\a_k$ and idiosyncratic volatilities $\s_k$ according to equation \eqref{wealthDistEq}. \citet{Fernholz:2002} and \citet{Fernholz/Koch:2016} show, respectively, that this is in fact true for the size distributions of total market capitalizations of U.S. stocks and total assets of U.S. financial intermediaries. One of this paper's contributions is to further demonstrate the validity of our econometric techniques using a new data set.

\begin{figure}[ht]
\begin{center}
\vspace{-10pt}
\hspace{-15pt}\scalebox{1.66}{ {\includegraphics[width=0.5\textwidth]{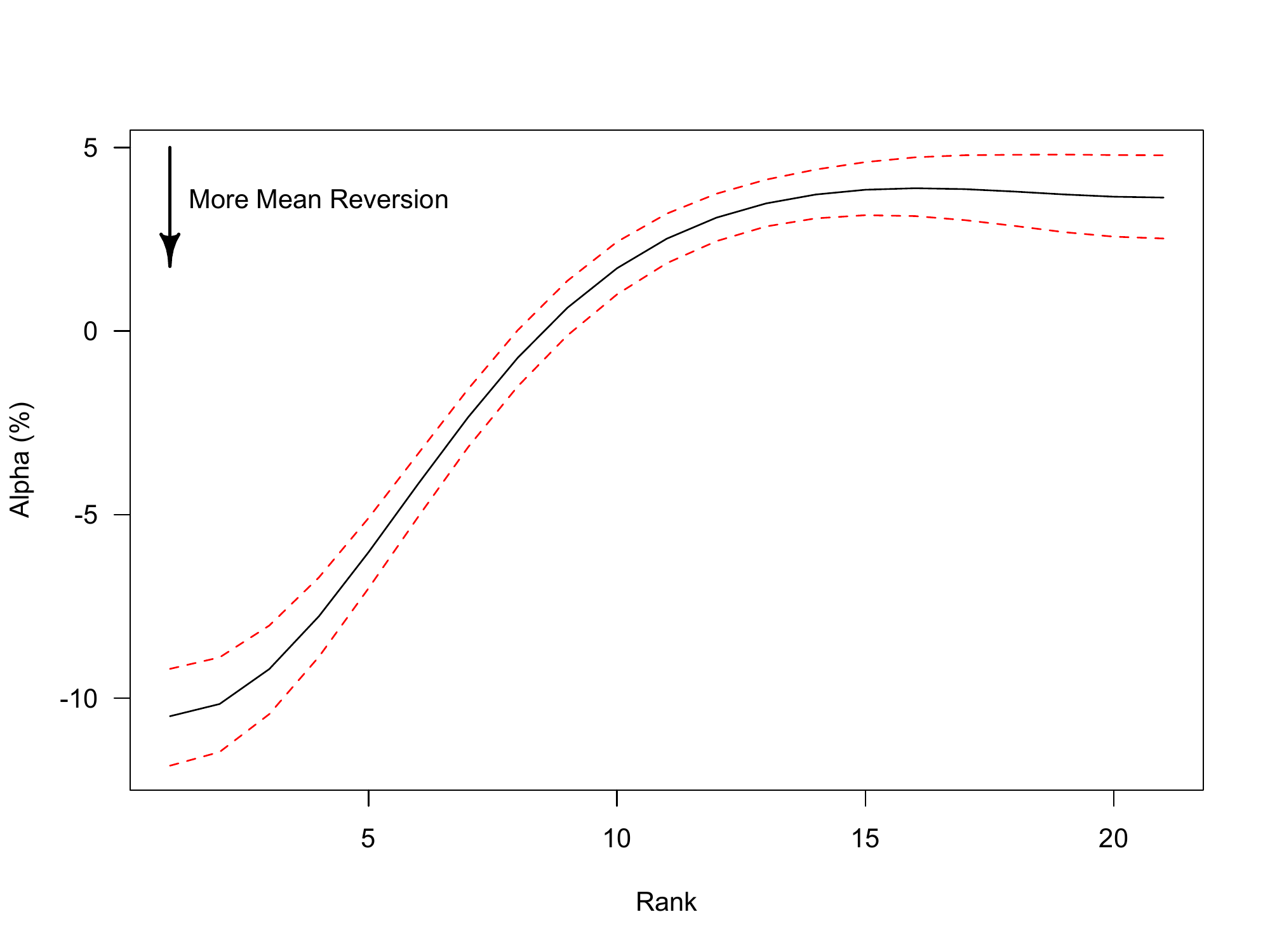}}}
\end{center}
\vspace{-24pt} \caption{Point estimates and 95\% confidence intervals of minus the reversion rates ($\a_k$) for different ranked commodities, 1980 - 2015.}
\label{alphasFig}
\end{figure}

The first step is to estimate the reversion rates $-\a_k$ for each rank $k = 1, \ldots, N$. As described in Section \ref{model}, these reversion rates measure the growth rates of different ranked commodity prices relative to the growth rate of all commodity prices together. In Figure \ref{alphasFig}, we plot annualized values of minus the reversion rates $\a_k$ for each rank in the distribution of relative normalized commodity prices together with 95\% confidence intervals based on the results of 10,000 bootstrap resample estimates.

\begin{figure}[ht]
\begin{center}
\vspace{-30pt}
\hspace{-15pt}\scalebox{1.65}{ {\includegraphics[width=0.5\textwidth]{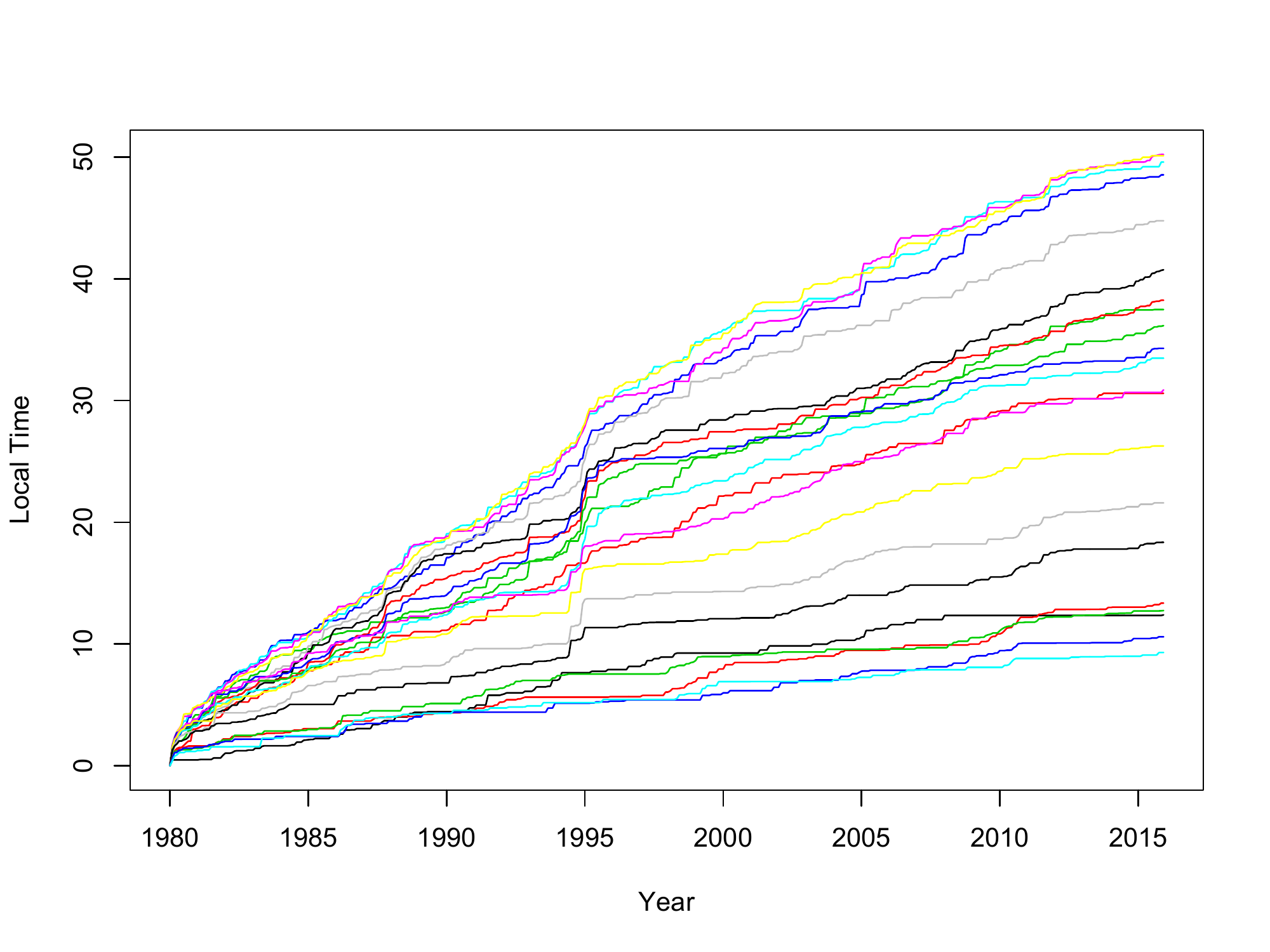}}}
\end{center}
\vspace{-24pt} \caption{Local time processes ($\L_{\log\tilde{x}_{(k)} - \log\tilde{x}_{(k+1)}}$) for different ranked commodities, 1980 - 2015.}
\label{localTimesFig}
\end{figure}

These parameters are estimated using the procedure described in Section \ref{estimation}. In particular, these estimated reversion rates are generated by first estimating the local time parameters $\k_k$ according to equations \eqref{estLocalTime3} and \eqref{estKappa}, and then generating estimates of the parameters $\a_k$ according to equation \eqref{alphaKappa} from Lemma \ref{localTimeAKLemma}. Figure \ref{localTimesFig} plots the evolution of the local time processes $\L_{\log\tilde{x}_{(k)} - \log\tilde{x}_{(k+1)}}$, $k = 1, \ldots, N-1$, which we use to construct our estimates of the reversion rates $-\a_k$.

The confidence intervals in Figure \ref{alphasFig} are generated using this same procedure, only with bootstrap resamples instead of the original full sample. These confidence intervals show that the deviations from Gibrat's law for commodity prices during the 1980 - 2015 period are highly statistically significant. This observation confirms the usefulness of our rank-based methods, since these methods allow for growth rates that vary across the distribution of relative commodity prices in the realistic manner shown in Figure~\ref{alphasFig}.

\begin{figure}[ht]
\begin{center}
\vspace{-10pt}
\hspace{-15pt}\scalebox{1.65}{ {\includegraphics[width=0.5\textwidth]{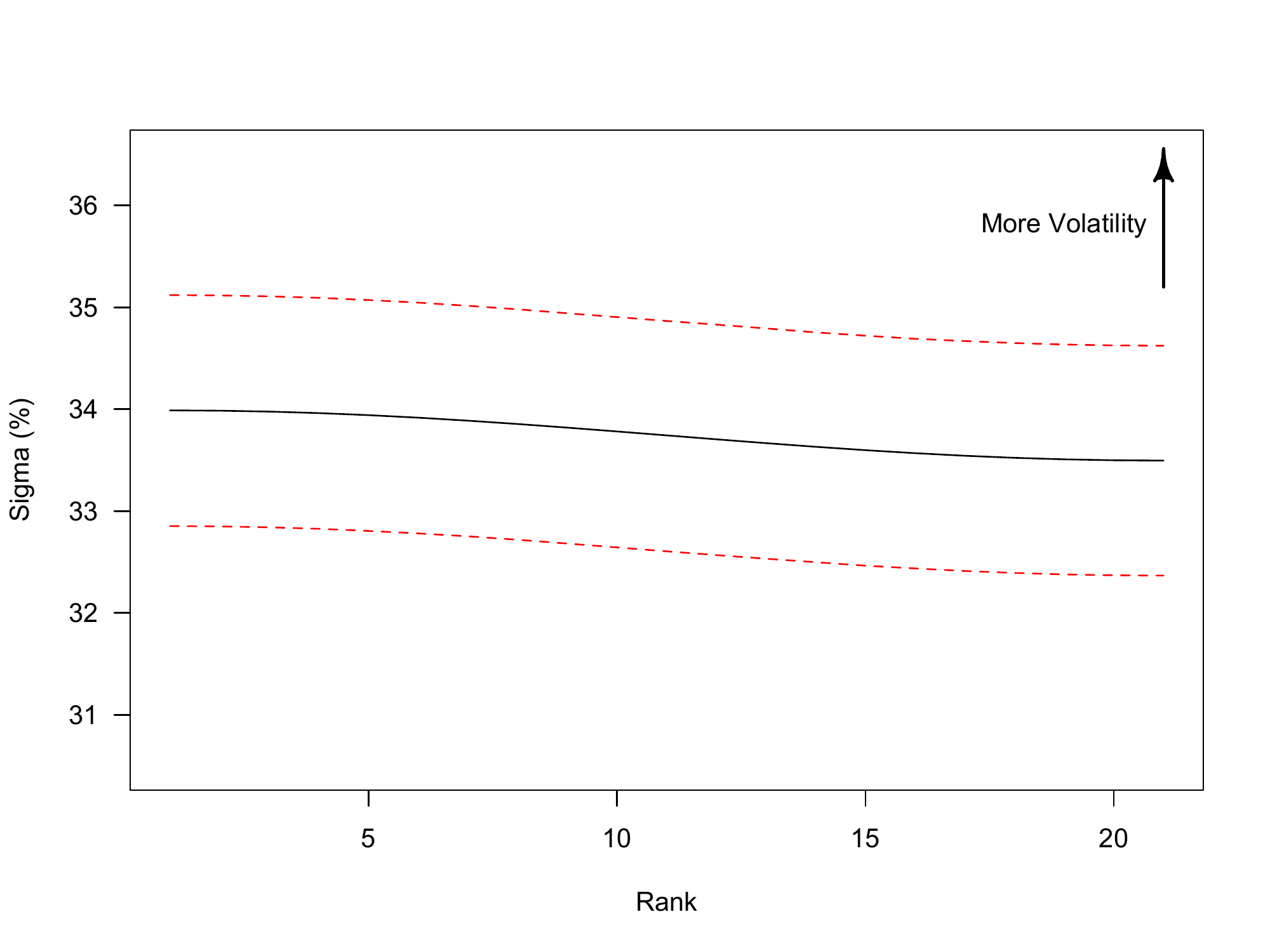}}}
\end{center}
\vspace{-24pt} \caption{Point estimates and 95\% confidence intervals of standard deviations of idiosyncratic commodity price volatilities ($\s_k$) for different ranked commodities, 1980 - 2015.}
\label{sigmasFig}
\end{figure}

The next step is to estimate the idiosyncratic volatilities $\s_k$, which is accomplished using the discrete-time approximation given by equation \eqref{estSigmaK}. Figure \ref{sigmasFig} plots annualized estimates for these parameter values for each rank in the distribution of relative normalized commodity prices together with 95\% confidence intervals based on the results of 10,000 bootstrap resample estimates. The estimates and confidence intervals for the parameters $\a_k$ and $\s_k$ in Figures \ref{alphasFig} and \ref{sigmasFig} are smoothed across different ranks using a Gaussian kernel smoother. Following \citet{Fernholz/Koch:2016}, we smooth these parameters between 1 and 100 times and then choose the number of smoothings within this range that minimizes the squared deviation between the predicted relative commodity prices according to equation \eqref{wealthDistEq} and the average observed relative commodity prices for the period 1981-2015.\footnote{The commodity prices are normalized to all equal each other at the start of our sample period in 1980. Since it takes a number of months for these initially equal relative prices to converge to a stationary distribution, we remove the first year of data when generating observed average relative prices to compare to predicted relative prices for the purposes of smoothing the parameters $\a_k$ and $\s_k$.}

\begin{figure}[ht]
\begin{center}
\vspace{-30pt}
\hspace{-15pt}\scalebox{1.65}{ {\includegraphics[width=0.5\textwidth]{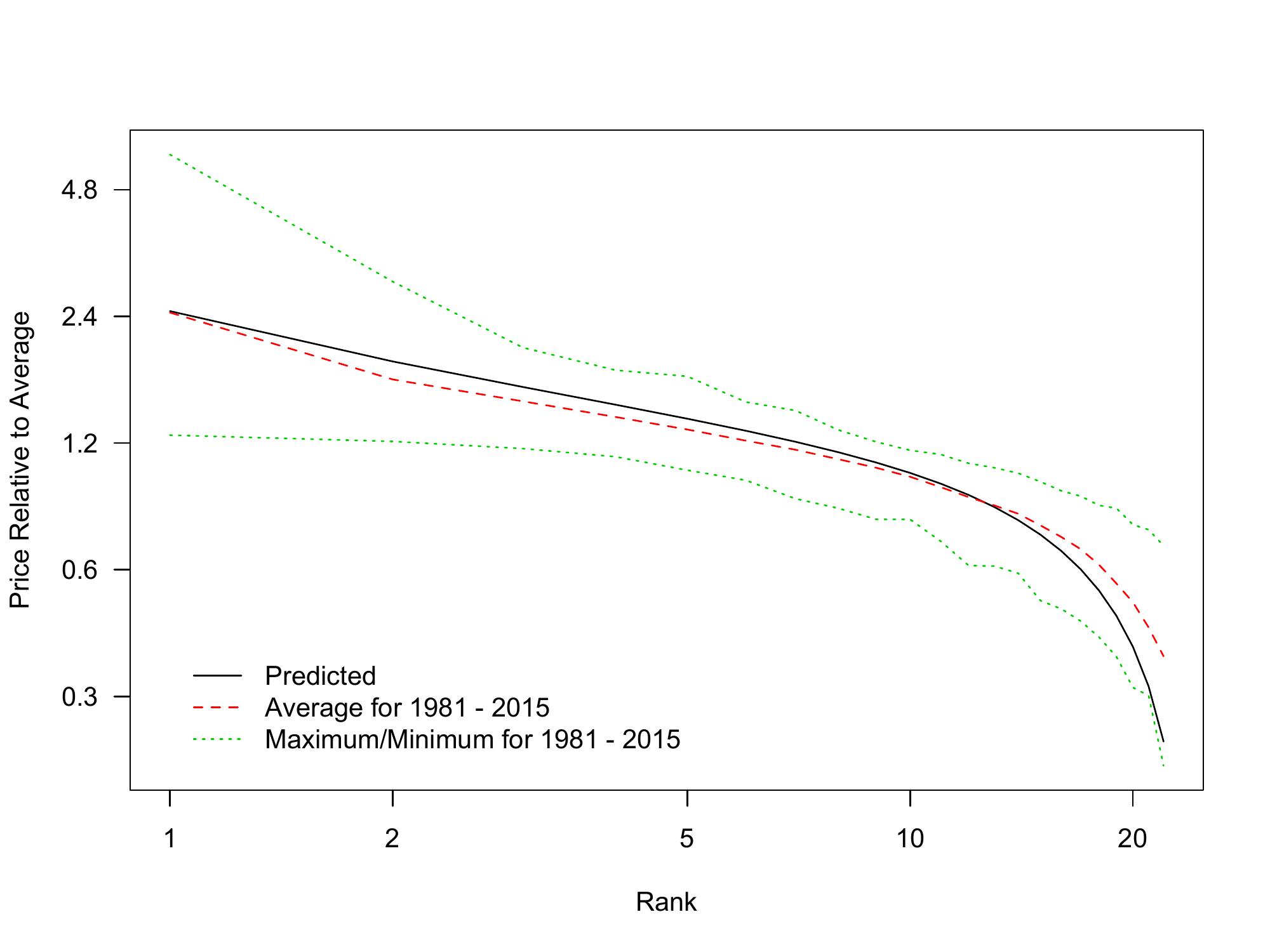}}}
\end{center}
\vspace{-24pt} \caption{Relative commodity prices for different ranked commodities for 1981 - 2015 as compared to the predicted relative prices.}
\label{predvDataFig}
\end{figure}

How well do the reversion rates $-\a_k$ and idiosyncratic volatilities $\s_k$ reported in Figures \ref{alphasFig} and \ref{sigmasFig} replicate the true distribution of relative commodity prices? Figure \ref{predvDataFig} shows that these estimated parameters generate predicted relative commodity prices according to equation \eqref{wealthDistEq} that do in fact match the average relative commodity prices observed during the 1980 - 2015 sample period. The squared deviation between predicted and observed average relative commodity prices over this sample period is 0.143. Thus, we further confirm the validity of our econometric methods using commodity prices data.


\subsection{A ``Size'' Effect for Commodities}
One of the implications of Theorem \ref{wealthDistThm} is that there is a stationary distribution of relative commodity prices if and only if $\a_1 + \cdots + \a_k < 0$. In other words, the growth rates of the prices of the higher-priced, higher-ranked commodities must on average be lower than the growth rates of the prices of the lower-priced, lower-ranked commodities, otherwise there is no stationary distribution of relative commodity prices. This necessary condition is essentially a mean-reversion condition.

Suppose that we interpret the processes $x_i$ in equation \eqref{wealthDynamics} from Section \ref{model} as the total market capitalizations of stocks. In this case, the dynamics of the processes $x_i$ correspond to capital gains, and hence the mean-reversion condition from Theorem \ref{wealthDistThm} implies that bigger stocks must generate smaller capital gains than smaller stocks. In other words, the mean-reversion condition from Theorem \ref{wealthDistThm} offers a structural, econometric explanation for the well-known size effect for stocks---the tendency for U.S.\ stocks with large total market capitalizations to generate lower average returns than U.S.\ stocks with small total market capitalizations \citep{Banz:1981,Fama/French:1993}. Indeed, this condition implies that a long-run size effect for capital gains is a necessary consequence of a stationary and realistic distribution of total stock market capitalizations.

\begin{figure}[ht]
\begin{center}
\vspace{-10pt}
\hspace{-15pt}\scalebox{1.66}{ {\includegraphics[width=0.5\textwidth]{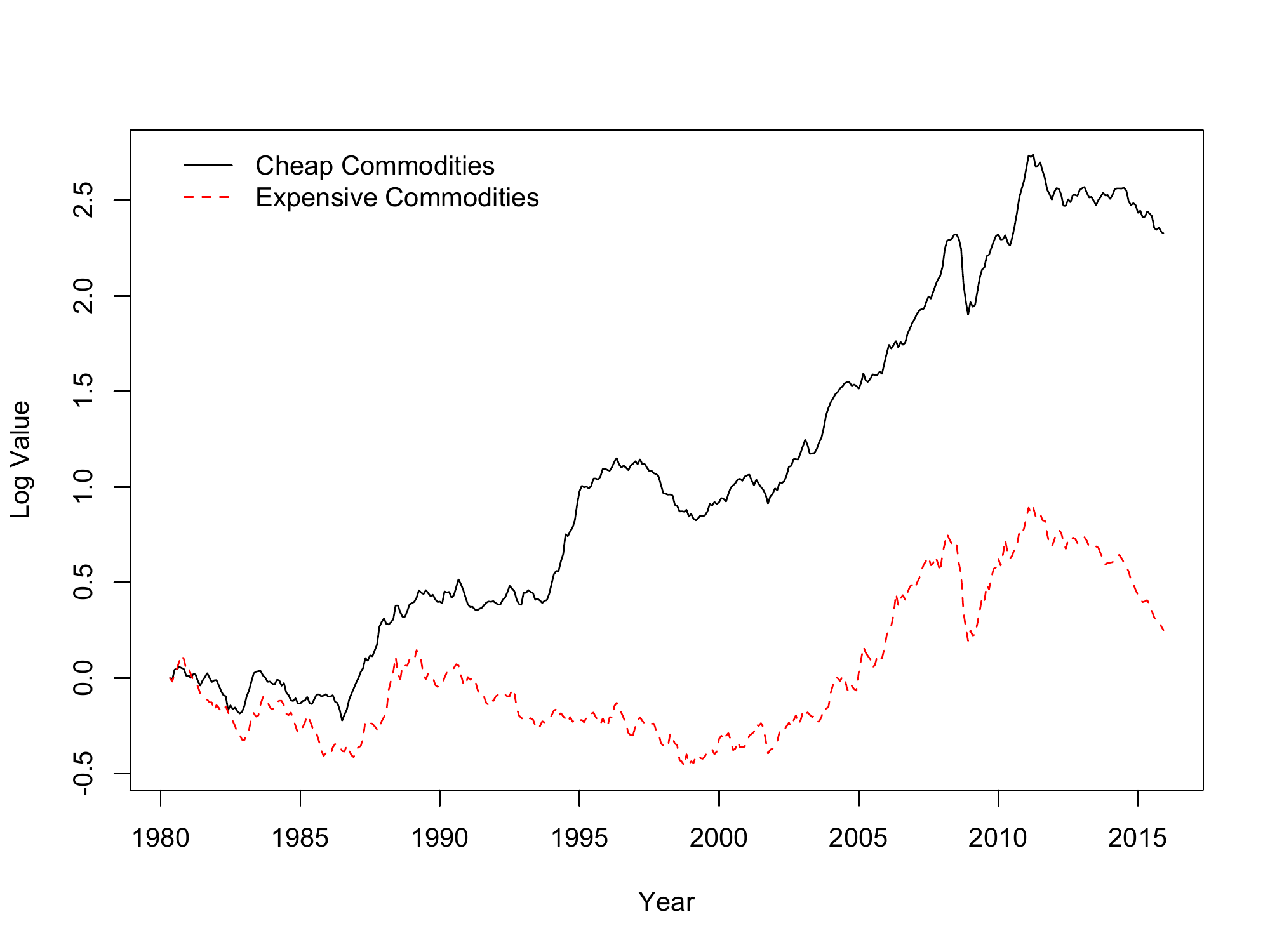}}}
\end{center}
\vspace{-24pt} \caption{Log returns for cheap-commodities and expensive-commodities portfolios, 1980 - 2015.}
\label{sizeFig1}
\end{figure}

This surprising implication of Theorem \ref{wealthDistThm} offers a testable prediction for our commodity prices data---there should be a generalized ``size'' effect for commodities in which higher-ranked, higher-priced, ``bigger,'' commodities generate lower returns on average than lower-ranked, lower-priced, ``smaller,'' commodities. In Figure \ref{sizeFig1}, we plot the log values over time of a portfolio that invests equal quantities in the eleven most expensive commodities in each month and a portfolio that invests equal quantities in the eleven cheapest commodities in each month. More precisely, in each month $t$, the expensive commodities portfolio is rebalanced to invest an equal quantity of the portfolio value in month $t$ in each of the eleven most expensive (highest ranked) commodities in month $t$. Conversely, in each month $t$, the cheap commodities portfolio is rebalanced to invest an equal quantity of the portfolio value in month $t$ in each of the eleven cheapest (least expensive, lowest ranked) commodities in month $t$. Figure \ref{sizeFig1} demonstrates a clear and large generalized size effect for commodities, just as predicted by our econometric results in Section \ref{model}.

\begin{figure}[ht]
\begin{center}
\vspace{-30pt}
\hspace{-15pt}\scalebox{1.66}{ {\includegraphics[width=0.5\textwidth]{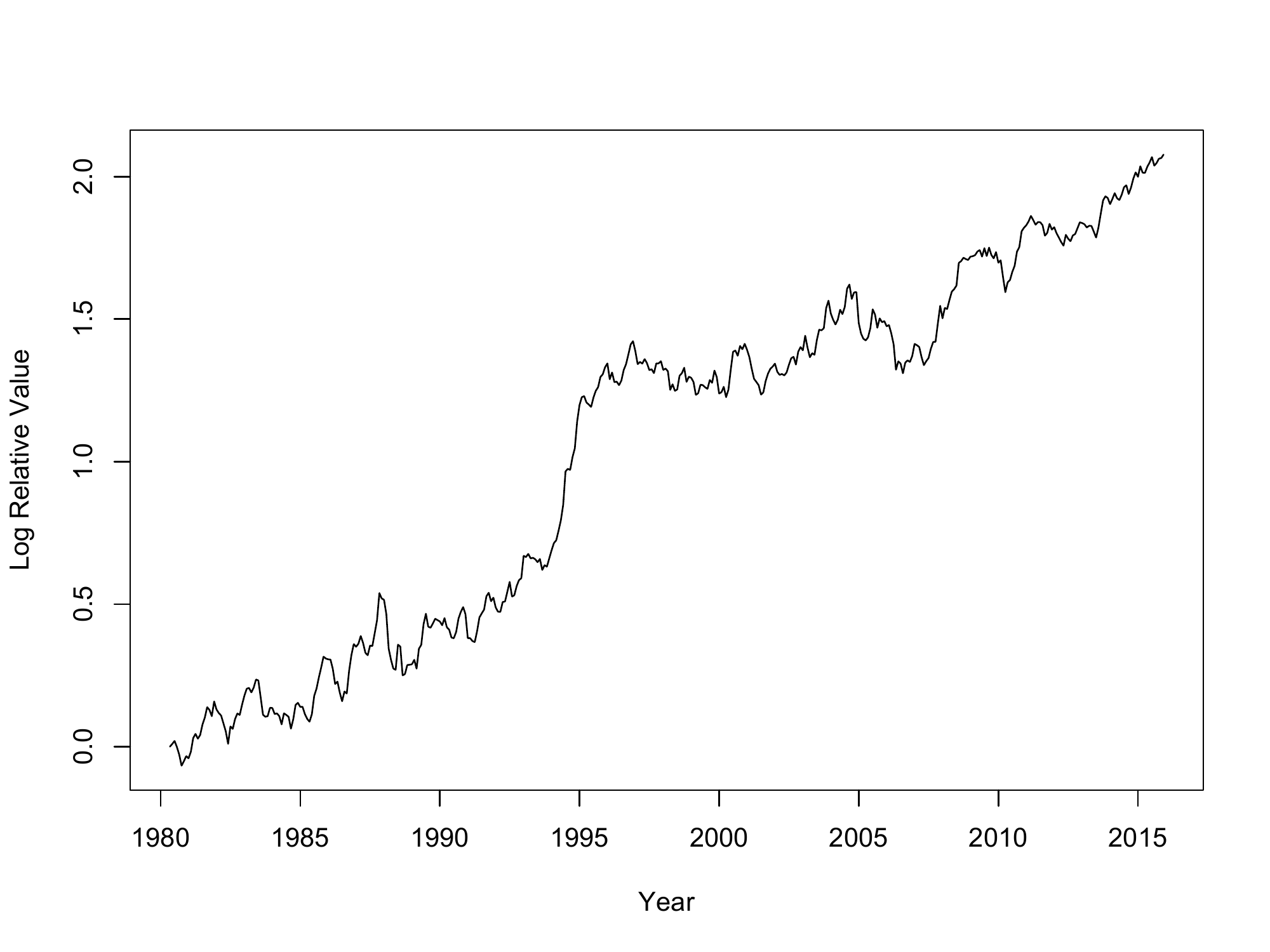}}}
\end{center}
\vspace{-24pt} \caption{Log return of cheap-commodities portfolio relative to expensive-commodities portfolio, 1980 - 2015.}
\label{sizeFig2}
\end{figure}

Figure \ref{sizeFig2} plots the log of the value of the cheap commodities portfolio relative to the value of the expensive commodities portfolio. This figure confirms the generalized size effect for commodities as in Figure \ref{sizeFig1}. In terms of standard percentage returns, the cheap commodities portfolio generates an average yearly (monthly) return of 8.62\% (0.59\%), while the expensive commodities portfolio generates an average yearly (monthly) return of 2.25\% (0.12\%).

Figure \ref{sizeFig2} also shows that the excess return of the cheap commodities portfolio relative to the expensive commodities portfolio does not appear to be highly positively correlated with either U.S.\ equity returns or the U.S.\ business cycle. This is surprising, since such positive correlations would be predicted from standard asset pricing theories \citep{Lucas:1978,Cochrane:2005}. Nonetheless, the generalized size effect for commodities predicted by the mean-reversion condition of Theorem \ref{wealthDistThm} and confirmed in Figures \ref{sizeFig1} and \ref{sizeFig2} is not necessarily inconsistent with standard equilibrium asset pricing theories.

The mean-reversion condition of Theorem \ref{wealthDistThm} implies that a generalized size effect for commodities is a necessary consequence of a realistic stationary distribution. It does not, however, imply anything about the properties of the excess returns from such a size effect. For example, the size effect shown in Figure \ref{sizeFig1} could be a reflection of greater risk for the portfolio of cheap commodities relative to the portfolio of expensive commodities, one possible explanation for the size effect among stocks \citep{Fama/French:1993}. It could also be a reflection of lower liquidity for the portfolio of cheap commodities relative to the portfolio of expensive commodities, another possible explanation for the size effect among stocks \citep{Acharya/Pederson:2005}. The mean-reversion condition of Theorem \ref{wealthDistThm} implies only that a generalized size effect for commodities is to be expected, regardless of whether or not such a size effect is a reflection of higher risk or lower liquidity.

It is beyond the scope of this paper to examine in detail the risk and liquidity properties of the two portfolio returns shown in Figure \ref{sizeFig1}, but such an analysis may yield interesting insight about the asset-pricing implications of our econometric results. Furthermore, although we have confirmed the existence of a new generalized size effect for commodities as predicted by Theorem \ref{wealthDistThm}, the generality of our nonparametric, rank-based econometric framework in Section \ref{model} suggests that generalized size effects should exist for other size and relative price distributions as well. As long as these other distributions are roughly stationary, then our theory predicts the existence of generalized size effects. Future research that attempts to uncover such new generalized size effects is likely to yield interesting conclusions.


\vskip 70pt

\section{Conclusion} \label{conclusion}

This paper presents rank-based, nonparametric methods that allow for the characterization of general power law distributions in random growth settings. We show that any stationary distribution in a random growth setting is shaped entirely by two factors---the idiosyncratic volatilities and reversion rates (a measure of cross-sectional mean reversion) for different ranks in the distribution. An increase in idiosyncratic volatilities increases concentration, while an increase in reversion rates decreases concentration. We also provide methods for estimating these two shaping factors using panel data.

Using data on a set of 22 global commodity prices from 1980 - 2015, we show that our rank-based, nonparametric methods accurately describe the distribution of relative normalized commodity prices. According to our econometric results, a necessary condition for the existence of a stationary distribution is that higher ranked (more expensive) commodity prices must grow more slowly than lower ranked (less expensive) commodity prices. In other words, our results predict a generalized ``size'' effect for commodities in which lower-priced commodities generate higher returns than higher-priced commodities. We confirm this prediction and show that a portfolio of lower-priced commodities has substantially higher returns than a portfolio of higher-priced commodities during the 1980 - 2015 period.

\vskip 70pt

\begin{spacing}{1.1}

\appendix
\section{Assumptions and Regularity Conditions} \label{assumptions}

In this appendix, we present the assumptions and regularity conditions that are necessary for the stable distribution characterization in Theorem \ref{wealthDistThm}. As discussed in Section \ref{model}, these assumptions admit a large class of continuous unit processes for the agents in our setup. The first assumption establishes basic integrability conditions that are common for both continuous semimartingales and It{\^o} processes.

\begin{ass} \label{basicAss}
For all $i = 1, \ldots, N$, the growth rate processes $\m_i$ satisfy
\begin{equation} \label{growthIntegrability}
 \intT|\m_i(t)|\,dt < \infty, \quad \text{$T > 0$,} \as,
\end{equation}
and the volatility processes $\d_{is}$ satisfy
\begin{align}
 & \intT\left(\d^2_{i1}(t) + \cdots + \d^2_{iM}(t)\right)\,dt < \infty, \quad \text{$T > 0$,} \as, \label{volIntegrability} \\
 & \d^2_{i1}(t) + \cdots + \d^2_{iM}(t) > 0, \quad \text{$t > 0$,} \as \label{volPositive} \\
 & \limt1\left(\d^2_{i1}(t) + \cdots + \d^2_{iM}(t)\right) \log\log t = 0, \label{volBounded} \as,
\end{align}
\end{ass}

Conditions \eqref{growthIntegrability} and \eqref{volIntegrability} are standard in the definition of an It{\^o} process, while condition \eqref{volPositive} ensures that agents' holdings of units contain a nonzero random component at all times. Condition \eqref{volBounded} is similar to a boundedness condition in that it ensures that the variance of agents' unit holdings does not diverge to infinity too rapidly.

The second assumption underlying our results establishes that no two agents' unit holdings be perfectly correlated over time. In other words, there must always be some idiosyncratic component to each agent's unit dynamics. Finally, we also assume that no agent's unit holdings relative to the total units for all agents shall disappear too rapidly.

\begin{ass} \label{corrAss}
The symmetric matrix $\r(t)$, given by $\r(t) = (\r_{ij}(t))$, where $1 \leq i, j \leq N$, is nonsingular for all $t > 0$, a.s.
\end{ass}

\begin{ass} \label{coherentAss}
For all $i = 1, \ldots, N$, the unit share processes $\theta_i$ satisfy
\begin{equation}
\limt1\log\theta_i(t) = 0, \as
\end{equation}
\end{ass}

\vskip 50pt

\section{Proofs} \label{proofs}

This appendix presents the proofs of Lemmas \ref{totalWealthLemma}, \ref{localTimeLemma}
\ref{localTimeAKLemma}, and \ref{estLocalTimeLemma}, and Theorem \ref{wealthDistThm}.

\begin{proofLemma1}
By definition, $x(t) = x_1(t) + \cdots + x_N(t)$ and for all $i = 1, \ldots, N$, $\theta_i(t) = x_i(t)/x(t)$. This implies that
\begin{equation*}
 dx(t) = \sum_{i=1}^Ndx_i(t) = \sum_{i=1}^N\theta_i(t)x(t)\frac{dx_i(t)}{x_i(t)},
\end{equation*}
from which it follows that
\begin{equation} \label{totalWealthDynamicsProof}
 \frac{dx(t)}{x(t)} = \sum_{i=1}^N\theta_i(t)\frac{dx_i(t)}{x_i(t)}.
\end{equation}
We wish to show that the process satisfying equation \eqref{totalWealthDynamics} also satisfies equation \eqref{totalWealthDynamicsProof}.

If we apply \ito's Lemma to the exponential function, then equation \eqref{totalWealthDynamics} yields
\begin{equation} \label{totalWealthDynamicsProof2}
\begin{aligned}
 dx(t) & = x(t)\m(t)\,dt + \frac{1}{2}x(t)\sum_{i,j=1}^N\theta_i(t)\theta_j(t)\left(\sum_{s=1}^M\d_{is}(t)\d_{js}(t)\right)\,dt \\
  & \qquad + x(t)\sum_{i=1}^N\sum_{s=1}^M\theta_i(t)\d_{is}(t)\,dB_s(t),
\end{aligned}
\end{equation}
a.s., where $\m(t)$ is given by equation \eqref{mu}. Using the definition of $\r_{ij}(t)$ from equation \eqref{rhoIJ}, we can simplify equation \eqref{totalWealthDynamicsProof} and write
\begin{equation} \label{totalWealthDynamicsProof3}
\frac{dx(t)}{x(t)} = \left(\m(t) + \frac{1}{2}\sum_{i,j=1}^N\theta_i(t)\theta_j(t)\r_{ij}(t)\right)\,dt + \sum_{i=1}^N\sum_{s=1}^M\theta_i(t)\d_{is}(t)\,dB_s(t).
\end{equation}
Similarly, the definition of $\m(t)$ from equation \eqref{mu} allows us to further simplify equation \eqref{totalWealthDynamicsProof3} and write
\begin{align}
\frac{dx(t)}{x(t)} & = \left(\sum_{i=1}^N\theta_i(t)\m_i(t) + \frac{1}{2}\sum_{i=1}^N\theta_i(t)\r_{ii}(t)\right)\,dt +\sum_{i=1}^N\sum_{s=1}^M\theta_i(t)\d_{is}(t)\,dB_s(t) \notag \\
& = \sum_{i=1}^N\theta_i(t)\left(\m_i(t) + \frac{1}{2}\r_{ii}(t)\right)\,dt + \sum_{i=1}^N\sum_{s=1}^M\theta_i(t)\d_{is}(t)\,dB_s(t).\label{totalWealthDynamicsProof4}
\end{align}

If we again apply \ito's Lemma to the exponential function, then equation \eqref{wealthDynamics} yields, a.s., for all $i = 1, \ldots, N$,
\begin{align}
 dx_i(t) & = x_i(t)\left(\m_i(t) + \frac{1}{2}\sum_{s=1}^M\d^2_{is}(t)\right)\,dt + x_i(t)\sum_{s=1}^M\d_{is}(t)\,dB_s(t) \notag \\
 & =  x_i(t)\left(\m_i(t) + \frac{1}{2}\r_{ii}(t)\right)\,dt + x_i(t)\sum_{s=1}^M\d_{is}(t)\,dB_s(t). \label{wealthDynamicsProof}
\end{align}
Substituting equation \eqref{wealthDynamicsProof} into equation \eqref{totalWealthDynamicsProof4} then yields
\begin{equation*}
 \frac{dx(t)}{x(t)} = \sum_{i=1}^N\theta_i(t)\frac{dx_i(t)}{x_i(t)},
\end{equation*}
which completes the proof.
\end{proofLemma1}

\begin{proofLemma2}
Agents' unit holding processes $x_i$ are absolutely continuous in the sense that the random signed measures $\m_i(t)\,dt$ and $\r_{ii}(t)\,dt$ are absolutely continuous with respect to Lebesgue measure. As a consequence, we can apply Lemma 4.1.7 and Proposition 4.1.11 from \citet{Fernholz:2002}, which yields equations \eqref{rankWealthDynamics} and \eqref{rankWealthShareDynamics1}.
\end{proofLemma2}

\begin{proofLemma3}
This relationship between the rank-based relative growth rate parameters $\a_k$ and the local time parameters $\k_k$ is established in the proof of Theorem \ref{wealthDistThm} below (see equation \eqref{alphaKProof} below). That proof also establishes the fact that $\a_N = -(\a_1 + \cdots + \a_{N - 1})$ (see equation \eqref{kappaAlphaProof} below).
\end{proofLemma3}

\begin{proofLemma4}
Consider the function $f_k(\theta_1, \ldots, \theta_N) = \theta_{(1)} + \cdots + \theta_{(k)}$, where $1 \leq k \leq N$. This function satisfies
\begin{equation*}
 \frac{\partial f_k}{\partial \theta_{l}} = 1,
\end{equation*}
for all $l = 1, \ldots, k$, and
\begin{equation*}
 \frac{\partial f_k}{\partial \theta_{l}} = 0,
\end{equation*}
for all $l = k+1, \ldots, N$. Furthermore, the support of the local time processes $\Lambda_{\log\theta_{(k)} - \log\theta_{(k+1)}}$ is the set $\{t \; : \; \theta_{(k)}(t) = \theta_{(k+1)}(t)\}$, for all $k = 1, \ldots, N - 1$. According to Theorem 4.2.1 and equations (3.1.1)-(3.1.2) of \citet{Fernholz:2002}, then, the function $f_k(\theta_1, \ldots, \theta_N) = \theta_{(1)} + \cdots + \theta_{(k)}$ satisfies the stochastic differential equation
\begin{equation} \label{localTimesProof}
\begin{aligned}
 d\log(x_{p_t(1)}(t) + \cdots + x_{p_t(k)}(t)) & - d\log x(t) = d\log f_k(\theta_1(t), \ldots, \theta_N(t)) \\
 & \qquad - \frac{\theta_{(k)}(t)}{2(\theta_{(1)}(t) + \cdots + \theta_{(k)}(t))}\,d\Lambda_{\log\theta_{(k)} - \log\theta_{(k+1)}}, \as,
\end{aligned}
\end{equation}
for all $k = 1, \ldots, N$.\footnote{Equation \eqref{localTimesProof} relies on the fact that $\log(x_{p_t(1)}(t) + \cdots + x_{p_t(k)}(t))$ is the value over time of a ``portfolio'' of unit holdings with weights of $\frac{\theta_{(l)}(t)}{\theta_{(1)} + \cdots + \theta_{(k)}}$ placed on each ranked unit holding $l = 1, \ldots, k$ and weights of zero placed on each ranked unit holding $l = k+1, \ldots, N$.} Equation \eqref{localTimesProof} is equivalent to
\begin{equation*}
\begin{aligned}
 d\log\left(\theta_{p_t(1)}(t) + \cdots + \theta_{p_t(k)}(t)\right) & = d\log \left(\theta_{(1)}(t) + \cdots + \theta_{(k)}(t)\right) \\
  & \qquad - \frac{\theta_{(k)}(t)}{2(\theta_{(1)}(t) + \cdots + \theta_{(k)}(t))}d\L_{\log\theta_{(k)} - \log\theta_{(k + 1)}}(t),
\end{aligned}
\end{equation*}
which confirms equation \eqref{estLocalTime1} from Lemma \ref{estLocalTimeLemma}.
\end{proofLemma4}

\begin{proofTheorem1}
This proof follows arguments from Chapter 5 of \citet{Fernholz:2002}. According to equation \eqref{rankWealthShareDynamics2}, for all $k = 1, \ldots, N$,
\begin{equation} \label{rankWealthShareProof1}
\begin{aligned}
 \log\theta_{(k)}(T) & = \intT\left(\m_{p_t(k)}(t) - \m(t)\right)\,dt +  \frac{1}{2}\L_{\log\theta_{(k)} - \log\theta_{(k + 1)}}(T) - \frac{1}{2}\L_{\log\theta_{(k - 1)} - \log\theta_{(k)}}(T) \\
 & \qquad + \sum_{s=1}^M\intT\d_{p_t(k)s}(t)\,dB_s(t) - \sum_{i=1}^N\sum_{s=1}^M\intT\theta_i(t)\d_{is}(t)\,dB_s(t).
\end{aligned}
\end{equation}
Consider the asymptotic behavior of the process $\log\theta_{(k)}$. Assuming that the limits from equation \eqref{kappa} exist, then according to the definition of $\a_k$ from equation \eqref{alphaK}, the asymptotic behavior of $\log\theta_{(k)}$ satisfies
\begin{equation} \label{rankWealthShareProof2}
\begin{aligned}
 \limT1\log\theta_{(k)}(T) & = \a_k + \frac{1}{2}\k_k - \frac{1}{2}\k_{k-1} + \limT1\sum_{s=1}^M\intT\d_{p_t(k)s}(t)\,dB_s(t) \\
 & \qquad  - \limT1\sum_{i=1}^N\sum_{s=1}^M\intT\theta_i(t)\d_{is}(t)\,dB_s(t), \as
\end{aligned}
\end{equation}
Assumption \ref{coherentAss} ensures that the term on the left-hand side of equation \eqref{rankWealthShareProof2} is equal to zero, while Assumption \ref{basicAss} ensures that the last two terms of the right-hand side of this equation are equal to zero as well (see Lemma 1.3.2 from \citealp{Fernholz:2002}). If we simplify equation \eqref{rankWealthShareProof2}, then, we have that
\begin{equation} \label{alphaKProof}
 \a_k = \frac{1}{2}\k_{k-1} - \frac{1}{2}\k_k,
\end{equation}
which implies that
\begin{equation} \label{alphaKappaProof}
 \a_k - \a_{k+1} = \frac{1}{2}\k_{k-1} - \k_k + \frac{1}{2}\k_{k+1},
\end{equation}
for all $k = 1, \ldots, N-1$. Since equation \eqref{alphaKProof} is valid for all $k = 1, \ldots, N$, this establishes a system of equations that we can solve for $\k_k$. Doing this yields the equality
\begin{equation} \label{kappaAlphaProof}
 \k_k = -2(\a_1 + \cdots + \a_k),
\end{equation}
for all $k = 1, \ldots, N$. Note that asymptotic stability ensures that $\a_1 + \cdots + \a_k < 0$ for all $k = 1, \ldots, N$, while the fact that $\a_N = \frac{1}{2}\k_{N-1} = -(\a_1 + \cdots + \a_{N-1})$ ensures that $\a_1 + \cdots + \a_N = 0$. Furthermore, if $\a_1 + \cdots + \a_k > 0$ for some $1 \leq k < N$, then equation \eqref{kappaAlphaProof} generates a contradiction since $\k_k \geq 0$ by definition. In this case, it must be that Assumption \ref{coherentAss} is violated and $\limT1\log\theta_{(k)}(T) \neq 0$ for some $1 \leq k \leq N$.

The last term on the right-hand side of equation \eqref{rankWealthShareDynamics3} is an absolutely continuous martingale, and hence can be represented as a stochastic integral with respect to Brownian motion $B(t)$.\footnote{This is a standard result for continuous-time stochastic processes \citep{Karatzas/Shreve:1991,Nielsen:1999}.} This fact, together with equation \eqref{alphaKappa} and the definitions of $\a_k$ and $\s_k$ from equations \eqref{alphaK}-\eqref{sigmaK}, motivates our use of the stable version of the process $\log\theta_{(k)} - \log\theta_{(k+1)}$. Recall that, by equation \eqref{stableVersion}, this stable version is given by
\begin{equation} \label{rankWealthShareDynamicsProof}
 d\left(\log\hat{\theta}_{(k)}(t) - \log\hat{\theta}_{(k+1)}(t)\right) = -\k_k\,dt + d\L_{\log\hat{\theta}_{(k)} - \log\hat{\theta}_{(k + 1)}}(t) + \s_k\,dB(t),
\end{equation}
for all $k = 1, \ldots, N-1$. According to \citet{Fernholz:2002}, Lemma 5.2.1, for all $k = 1, \ldots, N-1$, the time-averaged limit of this stable version satisfies
\begin{equation} \label{thetaKProof}
 \limT1\intT\left(\log\hat{\theta}_{(k)}(t) - \log\hat{\theta}_{(k+1)}(t)\right)\,dt = \frac{\s^2_k}{2\k_k} = \frac{\s^2_k}{-4(\a_1 + \cdots + \a_k)},
\end{equation}
a.s., where the last equality follows from equation \eqref{kappaAlphaProof}.

As shown by \citet{Banner/Fernholz/Karatzas:2005}, the processes $\log\hat{\theta}_{(k)} - \log\hat{\theta}_{(k+1)}$ are stationary if the condition $\a_1 + \cdots + \a_k < 0$ holds, for all $k = 1, \ldots, N$. Thus, by ergodicity, equation \eqref{wealthDistEq} follows from equation \eqref{thetaKProof}. To the extent that the stable version of $\log\theta_{(k)} - \log\theta_{(k+1)}$ from equation \eqref{rankWealthShareDynamicsProof} approximates the true version of this process from equation \eqref{rankWealthShareDynamics3}, then, the expected value of the true process $\log\theta_{(k)} - \log\theta_{(k+1)}$ will be approximated by $-\s^2_k/4(\a_1 + \cdots + \a_k)$, for all $k = 1, \ldots, N-1$.
\end{proofTheorem1}

\end{spacing}

\vfill

\begin{spacing}{1.2}

\bibliographystyle{chicago}
\bibliography{econ}

\end{spacing}

\vspace{200pt}

\end{document}

%% file: mathdef.tex
\usepackage{amsmath}
\usepackage{amstext}
\usepackage{amsbsy}
\usepackage{amsthm}
\usepackage{amsfonts}
\usepackage{amssymb}
\usepackage{amscd}
\usepackage{eucal}

\theoremstyle{plain}
\newtheorem{thm}{Theorem}[section]
\newtheorem{lem}[thm]{Lemma}

\newtheorem{ass}[thm]{Assumption}
\theoremstyle{definition}

\def\ito{It{\^o}}

\def\footnoterule{\hrule \kern2.6pt}

\def\intt{\int_0^t}
\def\intT{\int_0^T}

\def\rplu{[\,0,\infty)}

\def\as{\quad\text{\rm a.s.}}
\def\a{\alpha}
\def\b{\beta}

\def\L{\Lambda}
\def\r{\rho}
\def\s{\sigma}
\def\k{\kappa}
\def\d{\delta}

\def\O{\Omega}

\def\m{\mu}

\def\sgn{{\rm sgn}}
\def\empty{\varnothing}
\def\0T{[\,0,T]}

\def\limT#1{\lim_{T\to\infty}\frac{#1}{T}}
\def\limt#1{\lim_{t\to\infty}\frac{#1}{t}}

\def\F{\CMcal{F}}